\documentclass[pre,twocolumn,showpacs]{revtex4}
\usepackage{epsfig}
\usepackage{graphicx}
\usepackage{amsmath}
\usepackage{amssymb}

\begin{document}

\title{Surface modes and breathers in finite arrays of nonlinear waveguides}

\author{Yu. V. Bludov$^1$ and V. V. Konotop$^{1,2}$}
\affiliation{$^1$Centro de F\'{\i}sica Te\'{o}rica e Computacional, Universidade de Lisboa,
 Complexo Interdisciplinar, Avenida Professor Gama Pinto 2, Lisboa
1649-003, Portugal
\\
  $^2$ Departamento de F\'{\i}sica, Faculdade de Ci\^encias,
Universidade de Lisboa, Campo Grande, Ed. C8, Piso 6, Lisboa 1749-016, Portugal
%and Departamento de Matem\'aticas, E. T. S. de Ingenieros
%Industriales, Universidad de Castilla-La Mancha 13071 Ciudad Real, Spain
}


\begin{abstract}

We present the complete set of symmetric and antisymmetric ({\em edge} and {\em corner}) surface modes in finite one-- and two--dimensional arrays of
waveguides. We provide classification of the modes based on the anti-continuum limit, study their stability and bifurcations, and discuss relation
between surface and bulk modes. We put forward existence of {\em surface breathers}, which represent two-frequency modes localized about the array
edges.

\end{abstract}

\maketitle

\section{Introduction}

Waves at surfaces and interfaces
are known to
exhibit peculiar properties. Localized electronic states at a crystal edge, discovered by
Tamm~\cite{tamm}, were the first example of such phenomena. Later on it was found, that surfaces and interfaces are able to sustain localized waves
which  attracted a great deal of attention in different areas of physics, particularly due to  variety of practical applications,
like plasmonic waveguides~\cite{plas-wav}, sensors~\cite{sen1}, etc. Recently it was predicted theoretically~\cite{c-s2005} and observed
experimentally~\cite{c-s-exp}, that at the edge of a semi-infinite one-dimensional (1D) array of nonlinear waveguides there can exist discrete
surface solitons. Modes localized at finite distances from the edge were considered in~\cite{kiv-fd} and surface gap solitons between uniform media
and periodic lattice
were reported in~\cite{gap-teor}. Different type of surface modes in 2D arrays
were studied in \cite{2d-cr,VFMK}.

Structures with two surfaces give rise to novel properties of surface modes. For example, interaction of the two surface polaritons, supported by
each surface of a metallic film, results in creation of symmetric and antisymmetric modes~\cite{cam-mills}, which in their turn originate
polariton-assisted extraordinary transmittancy of the film~\cite{pol-trans}.

In this paper we describe discrete ({\em edge} and {\em corner}) surface modes in finite 1D and 2D arrays of nonlinear waveguides. We show that they
can be classified on the basis of the anti-continuum (AC) limit, similarly to the classification of intrinsic localized modes introduced in~\cite{vv}, and in this way the {\em complete} families of modes can be identified. We show that
surface modes can bifurcate either with other surface or with bulk modes and study the mode stability. We also report a new type of the surface
excitations --  {\em surface breathers} -- which represent two-frequency excitations localized in the vicinity of the array edges.

\section{Surface and bulk modes}

\subsection{The model and terminology}

We start with a finite array of $M$ nonlinear waveguides
described by the
discrete nonlinear Schr\"{o}dinger (DNLS)
equation~\cite{waveguide}
\begin{eqnarray}
\label{eq:mM} i\dot{q}_n+\sum_{n^\prime=1}^M(\delta_{n^\prime,n+1}+\delta_{n^\prime,n-1})q_{n^\prime}+\sigma\left|q_n\right|^2q_n=0.
\end{eqnarray}
Here $\dot{q}_n\equiv dq_n/d\zeta$, $\zeta$ is the propagation coordinate, $q_n$ is the dimensionless field amplitude in the $n$-th waveguide
($n=1,...,M$), $\sigma=1$ and $\sigma=-1$ stand for focusing and defocusing nonlinearities. We concentrate on modes having definite parity imposing
$q_n=q_{M+1-n}$ for symmetric and $q_n=-q_{M+1-n}$ for antisymmetric modes. Eq. (\ref{eq:mM}) possess two integrals of motion: the Hamiltonian
$H=-\sum_{n=1}^{M-1}\left(q^*_{n+1}q_n+q_{n+1}q_n^*\right)-\frac{\sigma}{2}\sum_{n=1}^{M}\left|q_n\right|^4$ and the total power
$P=\sum_{n=1}^{M}\left|q_n\right|^2$.

It worth to emphasize that the DNLS equation (\ref{eq:mM}) is a widely used model in the condensed matter physics~\cite{Scott} and in the theory of
Bose-Einstein condensates loaded in optical lattices~\cite{BK}, what makes the results reported below to be relevant  for a rather wide class of the
phenomena of the nonlinear physics of periodic and discrete structures.

Like in the well studied infinite case~\cite{reviews}, stationary modes of Eq.~(\ref{eq:mM}) are searched in the form
$q_n(\zeta)=Q_n\exp(-i\lambda \zeta)$, where $\lambda$ is the propagation constant, and the resulting equations are
\begin{eqnarray}
\label{eq:mMs} \lambda Q_n+\sum_{n^\prime=1}^M(\delta_{n^\prime,n+1}+\delta_{n^\prime,n-1})Q_{n^\prime}+\sigma Q_n^3=0.
\end{eqnarray}
The solutions for $\sigma=\pm 1$ are connected by the following symmetry reduction~\cite{vv}: if the $Q_n$ is a solution of (\ref{eq:mM}) for a
definite $\lambda$ and $\sigma=+1$, then $(-1)^nQ_n$ is a solution for $-\lambda$ and $\sigma=-1$.

In order to describe the whole diversity of solutions and to classify them~\cite{vv}, one has to consider the AC limit~\cite{ac}. To this end we rewrite Eq.~(\ref{eq:mM}) in terms of the rescaled stationary amplitudes
$v_n=Q_n/\sqrt{|\lambda|}$ as
\begin{eqnarray}
\label{eq:mM1} \sigma |\lambda| v_n(v_n^2+\sigma s)+\sum_{n^\prime=1}^M(\delta_{n^\prime,n+1}+\delta_{n^\prime,n-1})v_{n^\prime}=0,
\end{eqnarray}
where $s={\rm sign}(\lambda)$, and consider the limit $|\lambda| \to \infty$. In this limit $v_n$ become independent and for the case $\sigma s=-1$
acquire one of the three values: $v_n=-1$, $v_n=0$, and $v_n=+1$ ($1\le n\le M$). Thus, in the AC limit there exists
$N_{s}=\left(3^{\left[(M+1)/2\right]}-1\right)/2$ symmetric and $N_{a}=\left(3^{\left[M/2\right]}-1\right)/2$ antisymmetric modes (here square
brackets signify the integer part), and each mode can be coded~\cite{vv} by a sequence of $M$ symbols $-$, $0$, and $+$. The coding corresponds to
the limit of the infinite power: i.e., in particular, "0" does not refer to the zero intensity of a waveguide at a finite input intensity (see
Fig.~\ref{fig:fs}, and discussion below). As an example, an array of three waveguides has four symmetric:  $\{0+0\}$, $\{+++\}$, $\{+-+\}$, $\{+0+\}$
modes, and one antisymmetric mode $\{+0-\}$. A sequence, consisting of symbols $\{+,0,-\}$, is termed a "{\em word}", a word having only zeros is
referred to as {\em "empty"}, and a word having no $0$ (for example $\{+-+\}$) is called a {\em "simple"} word. Number of symbols in a word is called
the {\em length of the word}.

The first important property of the coding steams from the analytical continuation of the AC limit~\cite{ac} to $ |\lambda|>\lambda_{ac}$, where
$\lambda_{ac}$ is a constant (in the case of the infinite array $\lambda_{ac}\approx 5.4533$~\cite{vv}). This means that {\em all the words exhaust
all possible modes of a finite array} existing for $\lambda>\lambda_*$. Second, AC limit allows one to introduce a definition for a {\em surface
mode} as a {\em code consisting of two simple words separated by an empty word, the latter having the length not less than the lengths of each of the
simple words}. For example $\{+-+0000+-+\}$ is a surface mode of an array of $10$ waveguides. All other modes will be referred to as {\em bulk
modes}. The introduced terminology, being mathematically well defined, has relative physical meaning for finite $\lambda$: surface modes can
bifurcate with bulk modes, the both acquiring identical shapes in the bifurcation point.

\subsection{Bulk modes}

In the linear limit $P\to 0$ (or formally $\sigma\to 0$) the Eq.(\ref{eq:mMs}) possesses $M$ eigenvalues $\lambda_0^{(m)}=-2\cos{[\pi m/(M+1)]}$
corresponding to eigenmodes
\begin{equation}
Q_{0,n}^{(m)}=\sin\left(\frac{\pi n m}{M+1}\right), \quad m=1,...,M.
\label{eq:lin-modes}
\end{equation}

\begin{figure}[htb]
\centerline{\includegraphics{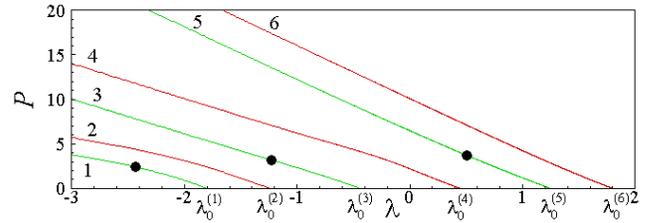}}
 \caption{Dependence of power $P$ \textit{vs} the propagation constant $\lambda$ for the bulk symmetric and antisymmetric modes in the case
 $\sigma=1$, $M=6$, which do not have the excitation threshold: 1 -- mode $\{00++00\}$, 2 -- mode $\{0+00-0\}$, 3 -- mode $\{+0--0+\}$, 4 -- mode $\{+0-+0-\}$,
 5 -- mode $\{+-++-+\}$, 6 -- mode $\{+-+-+-\}$. The bifurcation points with asymmetric modes are depicted by filled circles.} \label{fig:bulk}
\end{figure}

Thus one can expect that $M$ bulk modes have linear limit and thus do not possess an intensity threshold of excitation:
for these modes, when $\lambda\to\lambda_0^{(m)}$, the power $P^{(m)}$ of $m$-th mode ($m=1,\dots,M$) approaches zero
(see Fig.\ref{fig:bulk}). In order to determine the dependence $P^{(m)}(\lambda)$ near the linear limit we follow the
standard perturbation technique (see, e.g., \cite{bellman}), and look for a solution of (\ref{eq:mMs}) in a form of the
series
\begin{eqnarray}
Q_{n}=\epsilon Q_{0,n}^{(m)}+\epsilon^3 Q_{2,n}^{(m)}+ o(\epsilon^3),\\
\lambda=\lambda_0^{(m)}+\epsilon^2\lambda_2^{(m)}+o(\epsilon^2),
\end{eqnarray}
where we have introduced the small parameter $\epsilon=\sqrt{2P/(M+1)}\ll 1$.  Substituting the above expansions into Eq.(\ref{eq:mMs}) and gathering
the terms of the same order in $\epsilon$, we rewrite Eq.(\ref{eq:mMs}) in the form of a set of equations:
\begin{equation}
\lambda_0^{(m)} Q_{j,n}^{(m)}+\sum_{n^\prime=1}^M(\delta_{n^\prime,n+1}+\delta_{n^\prime,n-1})Q_{j,n^\prime}^{(m)}=
F_{j,n}^{(m)}.\label{eq:mul-scale}
\end{equation}
Here $F_{0,n}^{(m)}=0$, $F_{2,n}^{(m)}=-\lambda_2^{(m)} Q_{0,n}^{(m)}-\sigma\left(Q_{0,n}^{(m)}\right)^3$. As it is clear  Eq.(\ref{eq:mul-scale})
for $j=0$ describes a linear eigenmode and therefore is automatically satisfied, while considering solvability conditions for $j=2$  (it is
equivalent to orthogonality $F_{2,n}^{(m)}$ and $Q_{0,n}^{(m)}$, we obtain the corrections to the eigenvalues written in the form
\begin{equation}
 \lambda=\lambda_0^{(m)}-\epsilon^2 \sigma\frac{3+\delta_{m,(M+1)/2}}{4}.\label{eq:small-ampl}
\end{equation}

It follows from Eq.(\ref{eq:small-ampl}) that each of $M$ linear modes possesses its unique small-amplitude nonlinear analogue (from all diversity of
nonlinear modes only given $M$ bulk modes have small-amplitude solution).  This also proves that {\em no linear surface mode exists} (what
corroborates with the earlier findings for a semi-infinite array~\cite{kiv-fd}). Moreover, from Eq.(\ref{eq:small-ampl}) it follows, that in the
small-amplitude limit these modes are characterized by the linear dependence of the mode total power upon the propagation constant:
\begin{eqnarray}
P^{(m)}=\sigma(2M+2) \frac{\lambda_0^{(m)}-\lambda}{3+\delta_{m,(M+1)/2}}.\label{eq:linlim}
\end{eqnarray}
From Eq.(\ref{eq:linlim}) it is clear, that powers of these modes are decaying for $\sigma=1$ and increasing for
$\sigma=-1$ functions of $\lambda$. To single out the obtained modes in what follows they are referred to as {\em
quasi-linear} modes.

\subsection{Bifurcations of quasi-linear modes}

When the power increases one of two scenario of mode transformations is possible: either the branch of the quasi-linear
mode smoothly tends to a uniquely defined AC limit or at some $\lambda_*$ (alternatively $P_*$) it bifurcates giving
origin to some new solutions. As it is clear, each mode can bifurcate either with a mode of the same symmetry or with
an asymmetric mode (which in principle does not possess any symmetry, but in the bifurcation point acquires the given
symmetry). If the former even takes place then the quasi-linear mode is naturally classified by its AC limit. While
analytical description of other cases we leave for further studies, we mention that all numerical simulations we
performed with a finite number of waveguides have shown that no bifurcations of the quasi-linear modes with modes of
the same symmetry occurs: only asymmetric modes bifurcate from the quasi-linear ones. This allows us to use for the
latter modes the classification determined by their symmetric AC limit (i.e. by a word describing symmetric
ramification of the mode). Namely this notations are used in the figure captions and in the text whenever we speak
about quasi-linear modes.

To determine numerically the bifurcation points of the quasi-linear modes we consider the continuation of the mode by the parameter $\lambda$ (notice
that contrary the standard approach~\cite{ac,vv} now we "move" along the branch outwards the linear limit)
\begin{equation}
\frac{d\textbf{Q}}{d\lambda}=-\left[D_{Q}\textbf{F}\right]^{-1}\frac{\partial{\textbf{F}}(\textbf{Q},\lambda)}{\partial\lambda},
\end{equation}
where $\textbf{Q}={\rm col}\left(Q_n\right)$,
%\begin{widetext}
\begin{eqnarray}
&&\textbf{F}(\textbf{Q},\lambda)=\nonumber\\
&&{\rm col}\left(\lambda Q_n+\sum_{n^\prime=1}^M(\delta_{n^\prime,n+1}+\delta_{n^\prime,n-1})Q_{n^\prime}+\sigma Q_n^3\right),\nonumber
\end{eqnarray}
and the entries of the three-diagonal matrix $D_{Q}\textbf{F}$ are
\begin{equation}
(D_Q\textbf{F})_{n,n^\prime}=\delta_{n,n^\prime}\left(\lambda+3\sigma Q_n^2\right)+\delta_{n,n^\prime-1}+\delta_{n,n^\prime+1}.\label{eq:det}
\end{equation}
%\end{widetext}
This continuation of the quasi-linear modes is possible, while the matrix $D_{Q}\textbf{F}$ is invertible, i.e. its determinant is not equal to zero.
Thus the bifurcation point is determined by the equation $D\equiv D(\lambda)\equiv {\rm Det}|D_{Q}\textbf{F}|=0$.

As we already mentioned $M$ quasi-linear modes can bifurcate with asymmetric modes what is illustrated in
Fig.\ref{fig:bulk} obtained for $M=6$. There all symmetric modes possess additional bifurcation points, denoted by
filled circles. Although the consideration of modes without symmetry is beyond the scope of this paper, we now analyze
analytically the simplest case of  $M=2$ and $\sigma=1$ allowing the trivial solution. In that case there exists one
antisymmetric mode $Q^2_{1,2}=1-\lambda$ (with code $\{+-\}$) and one symmetric mode $Q^2_{1,2}=-1-\lambda$ (with code
$\{++\}$). Substituting the expressions for the field distribution of the antisymmetric mode $\{+-\}$ into
(\ref{eq:det}), we obtain, that $D(\lambda)=0$ only when $\lambda=\lambda_0^{(2)}=1$ (where the mode is born). For the
symmetric mode $\{++\}$ one verifies that similar to the previous case $D(\lambda)=0$ at the point of linear limit
$\lambda=\lambda_0^{(1)}=-1$, but also at $\lambda=\lambda_*=-2$, where the mode $\{++\}$ bifurcates with mode
$\{+0\}$. The mode $\{+0\}$ (for which $Q^2_{1,2}=-\lambda/2\pm\sqrt{\lambda^2/4-1}$) exists only when $\lambda \le
\lambda_*$, and at the point $\lambda_*$ its has the same field distribution as mode $\{++\}$ (so, we have a
pitchfork-type bifurcation).

\begin{figure}[ht]
%\centerline{\includegraphics[width=8.1cm]{fig0_3.ps}}
\centerline{\includegraphics{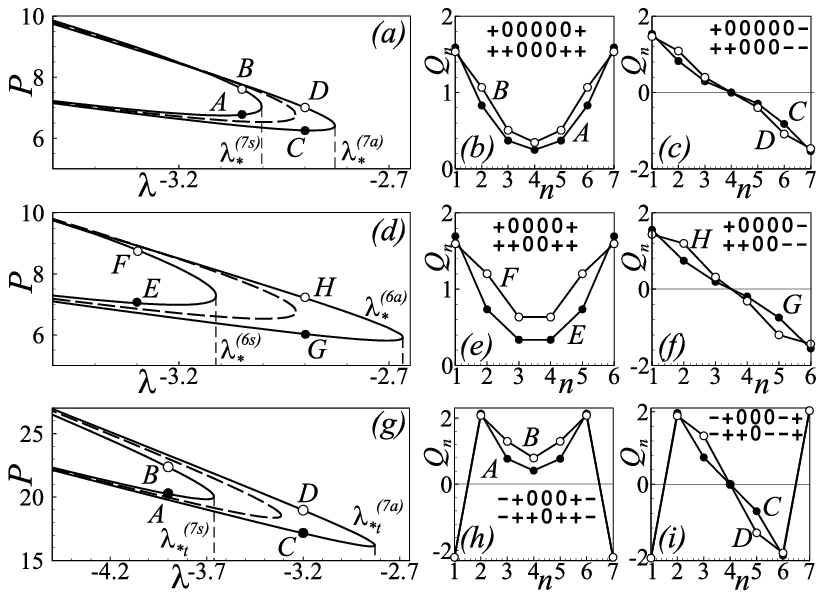}} \caption{$P$ {\it vs} $\lambda$ for fundamental (a), (d), and twisted (g) modes.
In panels b,c,e,f examples of the symmetric (A,B,E,F) and antisymmetric (C,D,G,H) fundamental surface modes, as well as
their classifications, are shown for $M=7$ (panels b,c) and $M=6$ (panels e,f). In panels h,i symmetric (A,B) and
antisymmetric (C,D) twisted surface modes for $M=7$ are shown. In all examples $\sigma=1$. For comparison, in panels
a,d, and g the dashed lines represent doubled powers of the same modes in a semi-infinite waveguide array.}
\label{fig:fs}
\end{figure}

\subsection{Surface modes in one-dimensional finite lattice}

Turning now to the analysis of surface modes, by analogy with an infinite array~\cite{Darm}, one can distinguish {\em
fundamental}  modes having in-phase distribution of the field near the edges and {\em twisted} surface modes, having
out-of-phase fields in the two waveguides bordering an edge. As an examples, in Fig.~\ref{fig:fs} we show the mode
patterns for arrays of $M=6$ and $M=7$ waveguides. All the modes shown require a threshold power to be excited. Two
different symmetric and antisymmetric fundamental surface modes   bifurcate with each other at $\lambda_*^{(Ms)}$ and
$\lambda_*^{(Ma)}$ (the codes of such modes are indicated in the respective panels of Fig.~\ref{fig:fs}). The pairs of
symmetric modes, (A,B) and (E,F), bifurcate in $\lambda_*^{(7s)}\approx -3.005$ and $\lambda_*^{(6s)}\approx -3.115$,
while the  bifurcation points of antisymmetric modes (C,D) and (G,H), are given by $\lambda_*^{(7a)}\approx -2.83$ and
$\lambda_*^{(6a)}\approx -2.67$. We observe that $\lambda_*^{(Ms)}<\lambda_*^{(Ma)}$ and thus the antisymmetric modes
are excited at lower field intensities. Comparing the bifurcation points of the modes with different $M$ we also
observe that for large enough arrays, i.e. at $M\to\infty$, the modes are transformed in the conventional surface modes
of a semi-infinite array. In this limit distinction between symmetric and antisymmetric modes disappears. Thus,
presence of two boundaries of a finite array essentially modifies a surface mode. The physical reason for this that in
a vicinity of the bifurcation point the modes are weakly localized near the array edges, and the field in waveguides
near the center of the array is non negligible (see Fig.\ref{fig:fs}). This leads to interaction between the modes,
supported by the two edges, what in its turn modifies the  patterns. At large values of $|\lambda|$ the modes are
strongly localized near the edges and interaction between them is weak, what results in the identical asymptotic
behavior of symmetric and antisymmetric modes in the AC limit clearly seen in Fig.~\ref{fig:fs} a,d.

The code of the twisted mode includes more nonzero symbols in comparison with the code of the fundamental mode. So,
since symbol "0" in the code of certain mode signifies zero field amplitude in the correspondent waveguide in the AC
limit, the power of twisted mode in AC limit should be higher than the power of fundamental mode. Nevertheless as shown
in Fig.~\ref{fig:fs} g, even in the vicinity of the bifurcation point twisted modes are excited at higher intensities
than the fundamental modes. Also due to higher field in the center of the waveguide array (in comparison with the
fundamental mode case) the interaction between two edges of the array is stronger, what is expressed by the relation
$\lambda_{*t}^{(7a)}-\lambda_{*t}^{(7s)}
> \lambda_*^{(7a)}-\lambda_*^{(7s)}$ (the bifurcation points are $\lambda_{*t}^{(7s)}\approx -3.665$ and
$\lambda_{*t}^{(7a)}\approx -2.83$). The main peculiarity of the diagram of the twisted modes is that surface modes A
and C bifurcate with the bulk modes B and D.

To study linear stability of the modes, we follow the standard steps analyzing the eigenvalue problem linearized about the mode, where $\beta$ is the
spectral parameter such that   ${\rm Im}(\beta)<0$ corresponds to a linearly unstable mode. The imaginary parts of $\beta$ for the fundamental modes
B, D, F, and H decrease with $\lambda \to -\infty$, i.e. the modes are unconditionally unstable (similar behavior was reported in~\cite{gen-sol}).
Symmetric fundamental modes A and E are unstable, but ${\rm Im}(\beta)$ increase with the decreasing of $\lambda$. The antisymmetric fundamental
modes C and G are unstable in the vicinity of the bifurcation point, but are stabilized at $\lambda<-3.13$ (for $M=7$) and $\lambda<-3.22$ (for
$M=6$). Stability analysis of the twisted modes shows, that symmetric modes A, B, and the antisymmetric mode D are unconditionally linearly unstable,
while the antisymmetric mode C is stable at $\lambda<-3.87$ (these results corroborate with the results of \cite{pel-kev} for the discrete modes of
infinite array).

\section{Surface breathers}

\begin{figure}[htb]
\centerline{\includegraphics{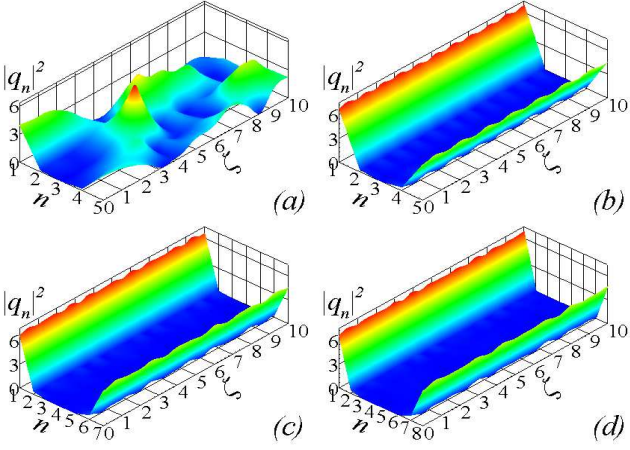}}
 \caption{(Color online) Surface breather for (a) $M=5$, $P=7$, $H=-14.175$; (b) $M=5$, $P=12$, $H=-37.602$; (c) $M=7$, $P=12.021$, $H=-37.463$ and (d) $M=8$, $P=12.022$, $H=-37.471$, excited by noise.} \label{fig:breath}
\end{figure}

Now we consider a novel type of localized modes -- surface breathers. Antisymmetric surface breathers can be constructed analytically for an array of
$M=5$ sites. In this case, Eq. (\ref{eq:mM}) possesses a solution of the following type (found using the known dynamics of a dimer~\cite{raghavan})
\begin{eqnarray}
z(\zeta)=\frac {4C}{P} \cdot\left\{
\begin{array}{ll}
{\rm cn}\left(C(\zeta-\zeta_0)/k,k\right), & 0<k<1 \\
 {\rm dn}\left(C(\zeta-\zeta_0),1/k\right), & k>1
\end{array}\right.,
\end{eqnarray}
where ${\rm cn}$ and ${\rm dn}$ are Jacobi elliptic functions, $z=2(|q_1|^2-|q_2|^2)/P$ is the intensity contrast, $k=C/(\sqrt{2}\varrho(P))$ is the
elliptic modulus,
$$\zeta_0=k F(\arccos(z(0)P/(4C)),k)/C$$
$F(\phi,k)$ is the incomplete elliptic integral of the first kind,
$$C^2=\varrho^2-H/2-P^2/16-2,$$ and $\varrho=(P^2/2+2H+4)^{1/4}$. When $k \gg 1$, the function
$$ z(\zeta)\approx
(4C/P)\left[1-\sin^2(C(\zeta-\zeta_0))/2k^2\right], $$
describes a mode concentrated near the edges $n=1,5$ (see Fig.~\ref{fig:breath}) and
oscillating with the period $2K(1/k)/C$ ($K(k)$ is the complete elliptic integral of the first kind). In the limit $k \to \infty$ (taken at a
constant power $P$) the Hamiltonian achieves it minimal value $H=-P^2/4-2$ and the intensity contrast becomes a constant. In that case the surface
breather transforms into a fundamental surface mode (e.g. the breathers in Fig.~\ref{fig:breath} a and b transform into the surface modes with $P=7$,
$\lambda=-3.5$, $H=-14.25$ and $P=12$, $\lambda=-6$, $H=-38$, respectively). Thus, a surface breather can be excited from the fundamental surface
mode by small detuning of its Hamiltonian from the minimum. We used this idea to construct the antisymmetric surface breathers for arrays of $M=7$
and $M=8$ waveguides (Fig.~\ref{fig:breath} c and d), which can not be constructed analytically. Notice that when the Hamiltonian possesses its
minimal value, the surface breathers of Fig.~\ref{fig:breath} c,d transform into stationary modes with $M=7$, $\lambda=-6$, $H=-38.185$ and $M=8$,
$\lambda=-6$, $H=-38.192$, respectively.

We tested the stability of breathers by direct numerical solution of differential Eq.~(\ref{eq:mM}), perturbing the initial profiles by noise with an
amplitude of order of 10\% of $Q_n$ in each waveguide. The breather, depicted in Fig.~\ref{fig:breath} a has shown unstable behaviour, while
breathers of Fig.~\ref{fig:breath} b-d demonstrated a stable one. Like in the case of the surface modes increasing of $|\lambda|$ results in
stabilization of a surface mode.

The case $M=4$ also allows for analytical construction of symmetric ($b=1$) and antisymmetric ($b=-1$) breathers. Now
$z=z_1+6f^\prime(z_1)/[24\wp(P(\zeta-\zeta_0)/4;g_2,g_3)-f^{\prime\prime}(z_1)]$, where $z_1$ is a root of the polynomial
$f(z)=a_4+a_3z+a_2z^2+a_1z^3-z^4$, $a_4=16(4-4H^2/P^2-H-P^2/16)/P^2$, $a_3=64b(H+P^2/8)/P^3$, $a_2=-8(10+2H+P^2/4)/P^2$, $a_1=8b/P$, and $\wp(x;g_2,g_3)$
is the Weierstrass elliptic function with $g_2=a_2^2/12-a_1a_3/4-a_4$ and $g_3=a_1a_2a_3/48+a_3^2-a_2^3/216-(a_1^2/16+a_2/6)a_4$.
The character of oscillations of the field along the waveguides is determined by $\Delta=g_2^3-27g_3^2$: for $\Delta \ne 0$ the solutions are oscillatory about a
nonzero average, but for $\Delta=0$  the solutions are aperiodic.

\section{Corner and edge modes}

Now we discuss surface modes of a 2D finite $M\times M$ array, where each waveguide is coupled with the nearest neighbors. The system is described by
coupled 2D DNLS equation
\begin{eqnarray}
i\dot{q}_{n,m}+
\sum_{m'=1}^{M}
\left(
\delta_{m',m+1}+\delta_{m',m-1}
\right)q_{n,m'}
\nonumber \\
+\sum_{n'=1}^{M}\left(
\delta_{n',n+1}+\delta_{n',n-1}\right)q_{n',m}
 + \sigma\left|q_{n,m}\right|^2q_{n,m}=0.
\label{eq:nm}
\end{eqnarray}
Now the total power is given by $P=\sum_{n,m=1}^{M}\left|q_{n,m}\right|^2$ and the symmetry reductions of the stationary modes
$q_{n,m}(\zeta)=Q_{n,m}\exp(-i\lambda \zeta)$ are as follows: if the $Q_{n,m}$ is a solution of (\ref{eq:nm}) for a definite $\lambda$ and
$\sigma=+1$, then $(-1)^{n+m}Q_{n,m}$ is a solution for $-\lambda$ and $\sigma=-1$.

\begin{figure}[htb]
\vspace{0.3cm}
%\centerline{
\includegraphics[height=3.2cm]{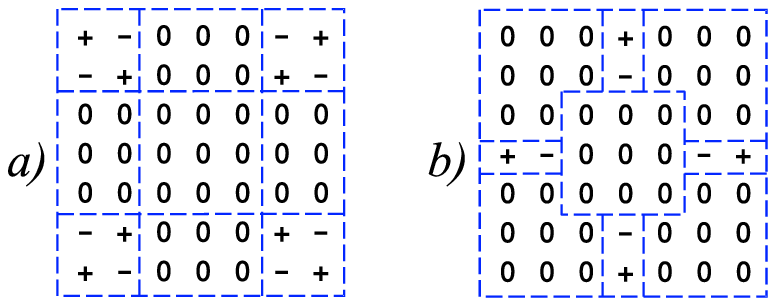}
%}
 \caption{ Examples of the fully symmetric corner (a) and edge (b) modes. The dashed lines outline empty and simple blocks.} \label{fig:def-cor-edge}
\end{figure}
Similar to the 1D case, each mode can be coded on a 2D {\em map} by an array of $M\times M$ symbols $-$, $0$, and $+$ (see examples in
Fig.~\ref{fig:def-cor-edge}), corresponding to its AC limit ($P\to\infty$). Splitting the array into blocks, we call {\em "empty"} a block consisting
of all zeros and "{\em simple}" a simply-connected block having no zeros. Now we define a {\em  corner} mode as a mode on a square array consisting
of only simple blocks at the corners, separated by empty blocks of higher dimensions (see an example in Fig.\ref{fig:def-cor-edge}a). Being
interested only in modes of a definite symmetry   we impose additional constrains: $q_{n,m}=q_{M+1-n,m}=q_{n,M+1-m}$ (for fully symmetric modes),
$q_{n,m}=-q_{M+1-n,m}=q_{n,M+1-m}$ (for symmetric-antisymmetric modes) and $q_{n,m}=-q_{M+1-n,m}=-q_{n,M+1-m}$ (for fully antisymmetric modes).
Similarly we identify an {\em edge} mode, whose code consists of simple blocks bordering edges of the array, but separated from each other and from
the corners by empty blocks (see the example in Fig.~\ref{fig:def-cor-edge}b). A symmetry constrains for the edge mode are as follows:
$q_{n,m}=q_{m,n}=q_{M+1-m,M+1-n}$ (for fully symmetric modes), $q_{n,m}=q_{m,n}=-q_{M+1-m,M+1-n}$ (for symmetric-antisymmetric modes) and
$q_{n,m}=-q_{m,n}=-q_{M+1-m,M+1-n}$ (for fully antisymmetric modes).

\begin{figure}[htb]
\centerline{\includegraphics{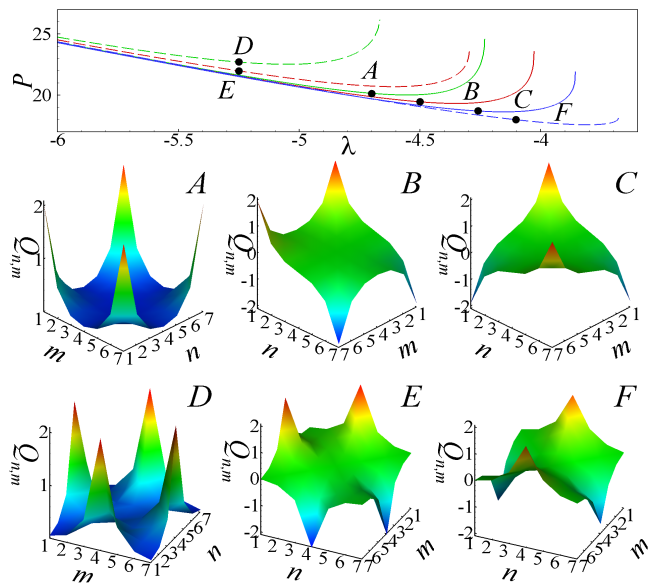}}
 \caption{(Color online) The total intensity $P$ {\it vs} the propagation constant
$\lambda$ (upper panel) for corner (solid lines) and edge (dashed lines) modes. Examples of the patterns of the corner (panels A-C) and edge
(panels D-F) modes are shown for $M=7$ and $\sigma=1$.  A and D  are fully symmetric modes, B and E are symmetric-antisymmetric modes, and C and F are fully antisymmetric modes.}
\label{fig:cor-edge}
\end{figure}

We restrict the consideration to the lowest-power fundamental modes, i.e. modes coded by one-site simple blocks. We
found that similar to the 1D case both corner and edge fully antisymmetric modes (C and F in Fig.~\ref{fig:cor-edge},
correspondingly) require lower threshold power of excitations that other types of the modes, while the fully symmetric
modes (A and D in Fig.~\ref{fig:cor-edge}, correspondingly) are excited at higher powers (see upper panel in
Fig.\ref{fig:cor-edge}). At the same time the distinction between the properties of the fully symmetric, semi-symmetric
and fully antisymmetric edge modes is stronger than the distinction between the properties of similar types of corner
modes. This occurs due to the smaller distance (and, hence, stronger interaction) between the excitation centered at
four edges in comparison with excitations centered at four corners (compare, e.g. field patterns for modes A and D, B
and E, C and F in Fig.\ref{fig:cor-edge}). The stability analysis shows that fully symmetric and
symmetric-antisymmetric modes A,B,D,E are unstable, but their instability increments ${\rm Im}(\beta)<0$ increase as
$\lambda$ decreases. The fully antisymmetric modes C and F are stabilized at $\lambda<-4.74$ and $\lambda<-5.4$,
respectively.

\section{Conclusion}

We have reported the complete families of (edge and corner) surface modes in arrays of 1D and 2D waveguides, presenting the classification exhausting
all possible stationary excitations. It has been shown that the surface modes belong to one-parametric branches of solutions, which bifurcate either
with other surface or with bulk modes, when the total power is properly changed. We have also found two-periodic modes, surface breathers, whose
intensity periodically oscillates staying localized about the edge of the array, and described a way of excitations of breathers starting with the
respective surface modes. The reported solutions, being well localized at the two (or four in the 2D case) surfaces and at the same time showing
nonzero intensity in the bulk of the array, can be of significant practical importance, by analogy with the polaritons assisting extraordinary
transmittancy~\cite{pol-trans}. Meantime we emphasize a number of open questions which were left unanswered by the present research. Among those we
mention thorough study of the linear stability of the surface breathers, mathematical justification of the complete classification of the
two-dimensional modes starting with the AC limit. These issue as well as specific practical outputs will be address elsewhere.

\acknowledgments

YVB. was supported by the FCT Grant No. SFRH/PD/20292/2004. VVK. acknowledges support of the Secretaria de Stado de Universidades e Investigaci\'{o}n
(Spain) under Grant No. SAB2005-0195. The work was supported by the FCT and European program FEDER (Grant No. POCI/FIS/ 56237/2004).


\begin{thebibliography}{10}
\newcommand{\enquote}[1]{``#1''}
\expandafter\ifx\csname url\endcsname\relax
  \def\url#1{\texttt{#1}}\fi
\expandafter\ifx\csname urlprefix\endcsname\relax\def\urlprefix{URL }\fi \providecommand{\eprint}[2][]{\url{#2}}

\bibitem{tamm}
I.~E. Tamm, Z. Phys. \textbf{76}, 849 (1932).

%\bibitem{maradudin}
%V.~M. Agranovich and D.~L. Mills, eds., \emph{Surface Polaritons.
%  Electromagnetic Waves at Surfaces and Interfaces} (North-Holland, Amsterdam,
%  1982).

\bibitem{plas-wav}
P.~Berini, R.~Charbonneau, N.~Lahoud, and G.~Mattiussi, J. Appl. Phys.
  \textbf{98}, 043109 (2005).

\bibitem{sen1} see e.g.
A.~G. Brolo, R.~Gordon, B.~Leathem, and K.~L. Kavanagh, Langmuir \textbf{20},
  4813 (2004).

\bibitem{c-s2005}
K.~G. Makris, {\it et al}
%S.~Suntsov, D.~N. Christodoulides, G.~I. Stegeman, and A.~Hache,
  Opt. Lett. \textbf{30}, 2466 (2005).

\bibitem{c-s-exp}
S.~Suntsov, {\it et al}
%K.~G. Makris, D.~N. Christodoulides, G.~I. Stegeman, A.~Hach\'{e},
%R.~Morandotti, H.~Yang, G.~Salamo, and M.~Sorel,
Phys. Rev. Lett.  \textbf{96}, 063901 (2006).

\bibitem{kiv-fd}
M.~I. Molina, R.~A. Vicencio, and Y.~S. Kivshar, Opt. Lett. \textbf{31},
  1693 (2006).

\bibitem{gap-teor} Y. V. Kartashov, V. A. Vysloukh, and L. Torner, Phys. Rev. lett.  \textbf{96},
073901 (2006).

\bibitem{2d-cr}
K.G. Makris,  J. Hudock, D.N. Christodoulides, G.I. Stegeman, O. Manela, and M. Segev,
Opt. Lett. \textbf{31}, 2774 (2006);
%
%\bibitem{KB}
H. Susanto,  P.G. Kevrekidis, B.A. Malomed, R. Carretero-Gonzalez, and D.J. Franzeskakis,
e-print nlin.PS/0607063.

\bibitem{VFMK}
R.~A. Vicencio, S. Flach, M.~I. Molina, and Y.~S. Kivshar, e-print cond-mat/0610049.

\bibitem{cam-mills}
R.~E. Camley and D.~L. Mills, Phys. Rev. B \textbf{29}, 1695 (1984).

\bibitem{pol-trans}
T.~W. Ebbesen, {\it et al}
%H.~J. Lezec, H.~F. Ghaemi, T.~Thio, and P.~A. Wolff,
Nature   \textbf{391}, 667 (1998).

\bibitem{vv}
G.L.~Alfimov, V.A.~Brazhnyi, and V.V.~Konotop, Physica D \textbf{194}, 127 (2004).

\bibitem{waveguide} D. N. Christodoulides and R. I. Joseph, Opt. Lett. \textbf{13},
794 (1988).

\bibitem{Scott} A. Scott, "{\em Nonlinear science: emergence of coherent structures.}" (Oxford University Press, 1999)

\bibitem{BK} V. A. Brazhnyi and V. V. Konotop, Mod. Phys. Lett. B {\bf 18}, 627
(2004).

\bibitem{reviews} see e.g. D. Hennig, and G. Tsironis, Phys. Rep. {\bf 307},
333 (1999);  P. G. Kevrekidis, K. \O . Rasmussen, and A. R. Bishop, {\it
Int. J. Mod. Phys.} {\bf B15}, 2833 (2001).

\bibitem{ac}
R.~S. MacKay and S.~Aubry, Nonlinearity \textbf{7}, 1623 (1994).

\bibitem{bellman} R. Bellman, "{\em Introduction to Matrix Analysis.}" (McGraw-Hill Education, 1970)

\bibitem{Darm} S. Darmanyan, A. Kobyakov, and F. Lederer, JETP {\bf 86}, 682 (1998)

\bibitem{gen-sol}
Y.~V. Kartashov, V.~A. Vysloukh, D.~Mihalache, and L.~Torner, Opt. Lett.
  \textbf{31}, 2329 (2006).

\bibitem{pel-kev}
D.E. Pelinovsky, P.G. Kevrekidis, D.J. Frantzeskakis, Physica D \textbf{212}, 1 (2005)

\bibitem{raghavan} V. M. Kenkre and D. K. Campbell, Phys. Rev. B {\bf 34}, 4959 (1986);
S.~Raghavan, A.~Smerzi, S.~Fantoni, and S.~R. Shenoy, Phys. Rev. A \textbf{59},
  620 (1999).

\end{thebibliography}
\end{document}